\documentclass{article}
\usepackage{graphicx,psfig,epsfig,amssym,rotate}
\def\RR{\Bbb R} 
\def\title{\begin{center}\Large\bf}
\def\author(s){\vspace{0.5cm}\large\rm}
\def\text{\end{center}}

\begin{document}

\title
Locating periodic orbits by Topological Degree theory

\author(s)
C.Polymilis$^a$, G. Servizi$^b$, Ch. Skokos$^{c,d}$, G. Turchetti$^b$ \&
M. N. Vrahatis$^e$
\footnote{E-mails: servizi@bo.infn.it (G.~S.), hskokos@cc.uoa.gr (Ch.~S.),
turchett@bo.infn.it (G.~T.), vrahatis@math.upatras.gr (M.~N.~V.)}

\vspace{0.3cm}

$^a${\it Department of Physics, University of Athens, Panepistimiopolis,
GR-15784 Zografos, Athens, Greece} \\
$^b${\it Department of Physics, Bologna University, Via Irnerio 46, I-40126 Bologna,
Italy and I.N.F.N. Sezione di Bologna, Via Irnerio 46, I-40126 Bologna, Italy}\\
$^c${\it Department of Mathematics, Division of Applied Analysis
and
Center for Research and Applications of Nonlinear Systems (CRANS),\\
University of Patras, GR-26500 Patras, Greece}\\
$^d${\it Research Center for Astronomy, Academy of Athens,
14 Anagnostopoulou str.,  GR-10673 Athens, Greece}\\
$^e${\it Department of Mathematics
and University of Patras Artificial Intelligence
 Research Center (UPAIRC),\\
University of Patras, GR-26110 Patras, Greece}\\

\text

\vspace{0.3cm}

\large

\section*{Abstract}
We consider methods based on the topological degree theory to compute
periodic orbits of area preserving maps. Numerical approximations of
the Kronecker integral and the application of Stenger's method allows us to 
compute the value of the topological degree in a bounded region of the
phase space. If the topological degree of an appropriate set of 
equations has a non--zero value, we know that there exists at least one
periodic orbit of a given period in the given region. We discuss in detail
the problems that these methods face, due to the existence of periodic
orbits near the domain's boundary and due to the discontinuity curves
that appear in maps defined on the torus. We use the characteristic bisection 
method for actually locating periodic orbits. We apply this method successfully,
both to the standard map, which is a map defined on the torus, and to the
beam--beam map which is a continuous map on the plane. Specifically we find
a large number of periodic orbits of periods up to 40, which  give us a clear 
picture of the dynamics of both maps.

\section{The topological degree (TD) and its computation}
We consider the problem of finding the solutions of a system of
nonlinear equations of the form
\begin{equation}
F_n(x) = \Theta_n, \label{eq:system}
\end{equation}
where $F_n = (f_1, f_2, \ldots, f_n) : D_n \subset \RR^n \rightarrow
\RR^n$ is a function from a domain $D_n$ into $\RR^n$, $\Theta_n = (0,
0,\ldots ,0)$
 and $x = (x_1, x_2, \ldots, x_n)$. The above system can be written as:
\begin{equation}
\begin{array}{lcl}
f_1(x_1, x_2, \ldots, x_n) & = & 0, \\
f_2(x_1, x_2, \ldots, x_n) & = & 0, \\
&\vdots& \\
f_3(x_1, x_2, \ldots, x_n) & = & 0.
\end{array}
\label{eq:sys}
\end{equation}

The topological degree (TD) theory gives us information on the existence
of solutions of the above system, their number and their nature
\cite{Kron,Pic92,Pic22,Cronin,Lloyd}. Kronecker
introduced the concept of the TD in 1869 \cite{Kron}, while Picard in
1892 \cite{Pic92} succeeded in
providing a theorem for computing the exact number of solutions 
of system (\ref{eq:system}).
Numerical methods based on the TD theory have been applied successfully
to numerous dynamical systems
(e.g. \cite{V95,V96etal,V97etal,V01etal,Kal,Per}).

In order to define the concept of the topological degree we consider the function
$F_n$ of system (\ref{eq:system})
to be continuous on the closure $\overline{D_n}$ of $D_n$, satisfying also
 $F_n(x) \neq \Theta_n$  for $x$ on the boundary $b(D_n)$ of
$D_n$. We also consider the solutions of (\ref{eq:system})  
to be simple
i.e.~the determinant of the corresponding Jacobian matrix
($J_{F_n}$) at the solution, to be different from zero. Then the {\it topological
degree (TD) of $F_n$ at $\Theta_n$ relative to $D_n$} is defined
as:
\begin{equation}
\mbox{deg}[F_n,D_n,\Theta_n] =
\sum_{x \in F_n^{-1} (\Theta_n)} \mbox{sgn}(\mbox{det} J_{F_n}(x)) =
N_+ - N_-,
\label{eq:TD}
\end{equation}
where det$J_{F_n}$ is the determinant of the Jacobian matrix of $F_n$,
sgn is the well-known sign function, $N_+$ the number of roots with
det$J_{F_n}>0$ and $N_-$ the number of roots with det$J_{F_n}<0$.
It is evident that if a nonzero value of deg$[F_n, D_n, \Theta_n]$ is obtained
then there exist at least one solution of system $F_n(x) = \Theta_n$ within $D_n$
\cite{Kron}.

A practical way to find the TD is the computation of Kronecker
integral \cite{Kron}. In particular,  under the assumptions of the
above--mentioned definition of the TD the deg$[F_n, D_n,
\Theta_n]$ can be computed as:
\begin{equation}
\mbox{deg}[F_n, D_n, \Theta_n] = \frac{\Gamma
(n/2)}{2\pi^{n/2}}\int\int\limits_{b(D_n)} \cdots \int
\frac{\sum_{i=1}^n A_i dx_i \ldots dx_{i-1} dx_{i+1} \ldots dx_n}
{(f_1^2+f_2^2+\ldots +f_n^2)^{n/2}}
\label{eq:Kron}
\end{equation}
where
\begin{equation}
A_i=(-1)^{n(i-1)} \left|
\begin{array}{ccccccc}
f_1 & \frac{\partial f_1}{\partial x_1} & \cdots & \frac{\partial
f_1}{\partial x_{i-1}} & \frac{\partial f_1}{\partial x_{i+1}} &
\cdots & \frac{\partial f_1}{\partial x_n} \\
f_2 & \frac{\partial f_2}{\partial x_1} & \cdots & \frac{\partial
f_2}{\partial x_{i-1}} & \frac{\partial f_2}{\partial x_{i+1}} &
\cdots & \frac{\partial f_2}{\partial x_n} \\
\vdots & \vdots & & \vdots & \vdots & & \vdots \\
f_n & \frac{\partial f_n}{\partial x_1} & \cdots & \frac{\partial
f_n}{\partial x_{i-1}} & \frac{\partial f_n}{\partial x_{i+1}} &
\cdots & \frac{\partial f_n}{\partial x_n}
\end{array}
\right|,
\label{eq:A}
\end{equation}
and $\Gamma(x)$ is the gamma function.

In order to find the number $N$ of solutions of system
(\ref{eq:system}) we consider the function
\begin{equation}
F_{n+1} = (f_1, f_2, \ldots, f_n, f_{n+1} ) : D_{n+1}\subset
\RR^{n+1} \rightarrow  \RR^{n+1}, \label{eq:Fexpand}
\end{equation}
where
\begin{equation}
f_{n+1} = y \;\mbox{det}J_{F_n},
\end{equation}
$ \RR^{n+1} : x_1, x_2, \ldots, x_n, y$ and $D_{n+1}$ is the product
of $D_n$ with a real interval on the $y$--axis containing $y=0$.
Then the exact number $N$ of the solutions of equation $F_n(x) =
\Theta_n$ is proven  to be \cite{Pic92}:
\begin{equation}
N= \mbox{deg}[F_{n+1}, D_{n+1}, \Theta_{n+1}]. \label{eq:Picard}
\end{equation}

By applying this result and using the computation of TD by Kronecker
integral (\ref{eq:Kron}) in the case of a set of two equations:
\begin{equation}
\begin{array}{lcl}
f_1(x_1, x_2) & = & 0, \\
f_2(x_1, x_2) & = & 0,
\end{array}
\label{eq:2eq}
\end{equation}
we find that the number $N$ of roots in the domain $D_2 =
[a,b]\times [c,d]$ is given by:
\begin{equation}
N=\frac{1}{2\pi} \int \limits_{b(D_2)} (P_1 dx_1 + P_2dx_2) +
\epsilon \int \int \limits_{D_2}
\frac{Qdx_1dx_2}{(f_1^2+f_2^2+\epsilon^2 J^2)^{3/2}},
\label{eq:N2}
\end{equation}
where $\epsilon$ is an arbitrary positive value, and
\begin{equation}
P_i=\frac{\left( f_1 \frac{\partial f_2}{\partial x_i} - f_2
\frac{\partial f_1}{\partial x_i} \right) \epsilon
J}{(f_1^2+f_2^2)(f_1^2+f_2^2+\epsilon^2 J^2)^{1/2}}, \;\; i=1,2
\;\;\;\;\;\; Q=\left|
\begin{array}{ccc}
f_1 & \frac{\partial f_1}{\partial x_1} & \frac{\partial f_1}{\partial x_2}\\
f_2 & \frac{\partial f_2}{\partial x_1} & \frac{\partial f_2}{\partial x_2}\\
J & \frac{\partial J}{\partial x_1} & \frac{\partial J}{\partial
x_2}
\end{array}
\right|, \label{eq:PQ}
\end{equation}
with $J$ denoting the determinant of the Jacobian matrix of
$F_2=(f_1, f_2)$.

Another method for computing the TD of $F_n$ at a domain $D_n$
is the application of Stenger's theorem \cite{Stenger,Mour}. Following this
approach for finding the TD, we only need to know the signs of functions $f_1$,
$f_2$, $\ldots$, $f_n$ in a `sufficient' set of points on the
boundary $b(D_n)$ of $D_n$.

\section{The characteristic bisection method}
The characteristic bisection method is based on the characteristic
polyhedron concept for the computation of roots of the equation
(\ref{eq:system}). The construction of a suitable
$n$--polyhedron, called the characteristic polyhedron, can be done
as follows. Let $M_n$ be the $2^n\times n$ matrix whose rows are
formed by all possible combinations of $-1$ and 1. Consider now an
oriented $n$--polyhedron $\Pi^n$, with vertices $V_k$,
$k=1,\ldots,2^n$. If the $2^n\times n$ matrix of signs associated
with $F$ and $\Pi^n$, whose entries are the vectors
\begin{equation}
\mbox{sgn}F_n(V_k) = (\mbox{sgn}f_1(V_k), \mbox{sgn}f_2(V_k),
\ldots, \mbox{sgn}f_n(V_k)),
\end{equation}
is identical to $M_n$, possibly after some permutations of these
rows, then $\Pi^n$ is called the characteristic polyhedron
relative to $F_n$. If $F_n$ is continuous, then, after some
suitable assumptions on the boundary of $\Pi^n$ we have: 
\begin{equation}
\mbox{deg}[F_n, \Pi^n, \Theta_n] = \pm 1 \neq 0.
\end{equation}
This means that there is at least one solution of  system
$F_n(x) = \Theta_n$ within $\Pi^n$.

\begin{figure}
\centerline{\epsfxsize=12.0cm \epsfbox{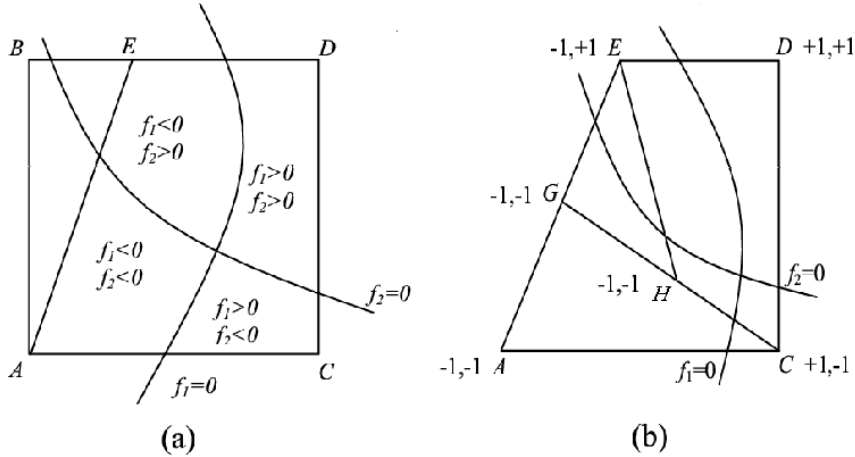}} 
\caption {(a) The polyhedron ABDC is noncharacteristic while the
polyhedron AEDC is characteristic. (b) Application of the
characteristic bisection method to the characteristic polyhedron
AEDC, giving rise to the polyhedra GEDC and HEDC, which are also
characteristic. } \label{f1}
\end{figure}

To clarify the characteristic polyhedron concept we consider a
function $F_2=(f_1, f_2)$. Each function $f_i$, $i=1,2$, separates
the space into a number of different regions, according to its
sign, for some regions $f_i<0$ and for the rest $f_i>0$, $i=1,2$.
Thus, in figure \ref{f1}(a) we distinguish between the regions
where $f_1<0$ and $f_2<0$, $f_1<0$ and $f_2>0$, $f_1>0$ and
$f_2>0$, $f_1>0$ and $f_2<0$. Clearly, the following combinations
of signs are possible: $(-,-)$, $(-,+)$, $(+,+)$ and $(+,-)$. Picking a
point, close to the solution, from each region we construct a
characteristic polyhedron. In this figure we can perceive a
characteristic and a noncharacteristic polyhedron $\Pi^2$. For a
polyhedron $\Pi^2$ to be characteristic all the above combinations
of signs must appear at its vertices. Based on this criterion,
polyhedron ABDC is not a characteristic polyhedron,
whereas AEDC is.
 A characteristic polyhedron can be considered
as a translation of Poincar\'{e}--Miranda hypercube \cite{V89}.

Next, we describe the characteristic bisection method. This method
simply amounts to constructing another refined characteristic
polyhedron, by bisecting a known one, say $\Pi^n$, in order to
determine the solution with the desired accuracy. We compute the
midpoint $M$ of an one-dimensional edge of $\Pi^n$, e.g. $\langle
V_i,V_j\rangle$. The endpoints of this one-dimensional line
segment are vertices of $\Pi^n$, for which the corresponding
coordinates of the vectors, sgn$F_n(V_i)$ and sgn$F_n(V_j)$ differ
from each other only in one entry. To obtain another
characteristic polyhedron  we compare the sign of $F_n(M)$ with
that of $F_n(V_i)$ and $F_n(V_j)$ and substitute $M$ for that
vertex for which the signs are identical. Subsequently, we reapply
the aforementioned technique to a different edge (for details we
refer the reader to \cite{V88a,V88b,V95,Per}).

To fully comprehend the characteristic bisection method we
illustrate in figure \ref{f1}(b) its repetitive operation on a
characteristic polyhedron $\Pi^2$. Starting from the edge AE we
find its midpoint G and then calculate its vector of signs, which
is $(-1,-1)$. Thus, vertex G replaces A and the new refined
polyhedron GEDC, is also characteristic. Applying the same
procedure, we further refine the polyhedron by considering the
midpoint H of GC and checking the vector of signs at this point.
In this case, its vector of signs is $(-1,-1)$, so that vertex G can
be replaced by vertex H. Consequently, the new refined polyhedron
HEDC is also characteristic. This procedure continues up to the
point that the midpoint of the longest diagonal of the refined
polyhedron approximates the root within a predetermined accuracy.

\section{Applications}

We apply methods based on the topological degree theory to compute
periodic orbits of two area preserving maps, the Standard map (SM)
\cite{Chir}, which is a map defined on the torus:
\begin{equation}
\begin{array}{lcl}
x' & = & x+y - \frac{k}{2 \pi} \sin (2 \pi x) \\
y' & = & y - \frac{k}{2 \pi} \sin (2 \pi x)
\end{array}
\;\; , \; \; \mbox{mod} (1), \;\; x,y \in [-0.5, 0.5),
\label{eq:SM}
\end{equation}
and the beam--beam map (BM) \cite{Baz,Pol}, which is a map defined
on $\RR^2$ :
\begin{equation}
\begin{array}{lcl}
x' & = & x \cos (2 \pi \omega) +( y +1- e^{-x^2})  \sin (2 \pi \omega), \\
y' & = & -x \sin (2 \pi \omega) +( y +1- e^{-x^2})  \cos (2 \pi
\omega)
.
\end{array}
\label{eq:BM}
\end{equation}

Given a dynamical map $M: \{x'=g_1(x,y), y'=g_2(x,y)\}$, the
periodic points of period $p$ are fixed points of the $p$--iteration
$M^p$ of the map and the zeroes of the function:
\begin{equation}
F= M^p - I = \left\{
\begin{array}{lcl}
f_1 & = & g_1^p(x,y)-x \\
f_2 & = & g_2^p(x,y)-y
\end{array}
\right. , \label{eq:Mp-I}
\end{equation}
where $I$ is the identity. 

\begin{figure}
\centerline{\epsfxsize=7.0cm \epsfbox{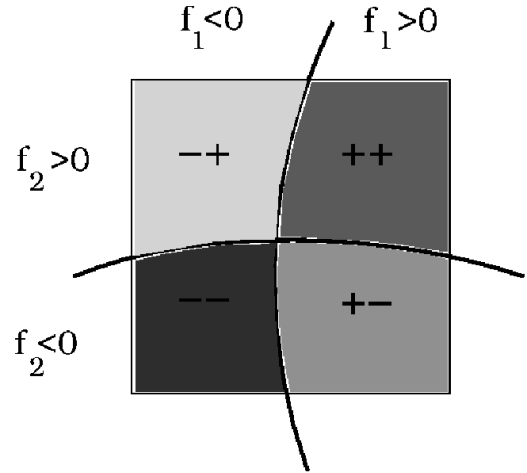}} 
\caption { Sketch of the domains where functions $f_1$ and $f_2$
of system (\ref{eq:Mp-I}) have a definite sign.} \label{f2}
\end{figure}

\begin{figure}
\centerline{\epsfxsize=14.0cm \epsfbox{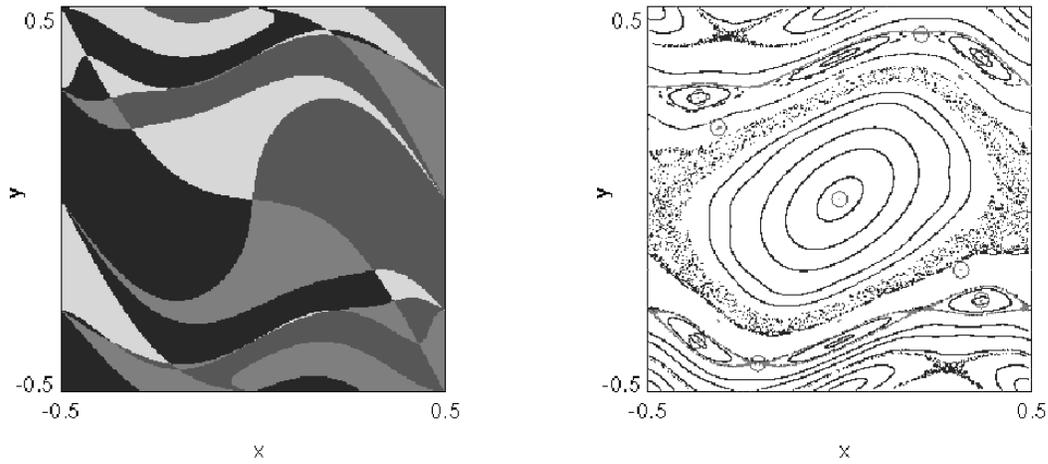}}
\caption
{Standard map (\ref{eq:SM}) for $k=0.9$: color map for $p=3$
iterations of the map computed on a square of $m\times m$ points
for $m=512$ (left panel); phase plot of the map (right panel). The
gray circles denote the positions of the zeros of the corresponding
function (\ref{eq:Mp-I}).} \label{fSM}
\end{figure}

\begin{figure}
\centerline{\epsfxsize=14.0cm \epsfbox{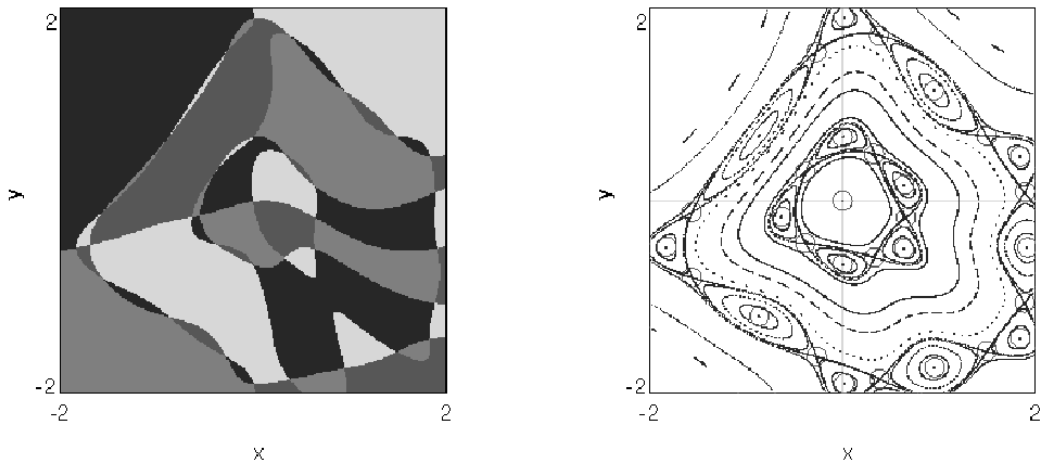}} \caption
{Beam-beam map (\ref{eq:BM}) for $\omega=0.21$: color map for
$p=5$ iterations of the map computed on a square of $m\times m$
points for $m=512$ (left panel); phase plot of the map (right
panel). The gray circles denote the positions of the zeros of the
corresponding function (\ref{eq:Mp-I}).} \label{fBM}
\end{figure}

One can use a color map to inspect the geometry of function $F$
(\ref{eq:Mp-I}) and to locate its zeroes. The color map is created by
choosing a lattice of $m\times m$ points and by associating to
each point a color chosen according to the signs of the functions
$f_1$, $f_2$ as shown in figure \ref{f2}. A simple algorithm
allows to detect the cells, formed by the lattice of $m\times m$
points, whose vertices have different colors. A cell is a
candidate to have a zero in its interior if the corresponding
topological degree is found to be different from zero. In figures
\ref{fSM} and \ref{fBM} we construct the color map and apply the
above mentioned algorithm for locating periodic orbits of period 3
for the SM and of period 5 for the BM, respectively. In both
figures the gray circles at the right panels 
denote the positions of the found periodic
orbits. We see that for both maps some periodic orbits were not
found because some of the four color domains close to the fixed
point were very thin. On the other hand, due to the discontinuity
of function $F$ (\ref{eq:Mp-I}), 
some zeros that do not correspond to real periodic orbits
were found for the SM (right panel of figure \ref{fSM}).

For maps defined on the torus like the SM (\ref{eq:SM}), the
computation of the TD using Stenger's method or the Kronecker
integral (\ref{eq:N2}) faces problems due to the presence of
discontinuity curves. Indeed Kronecker integral is defined on a
domain where the function $F$ (\ref{eq:Mp-I}) is continuous.

\begin{figure}
\centerline{\epsfxsize=14.0cm \epsfbox{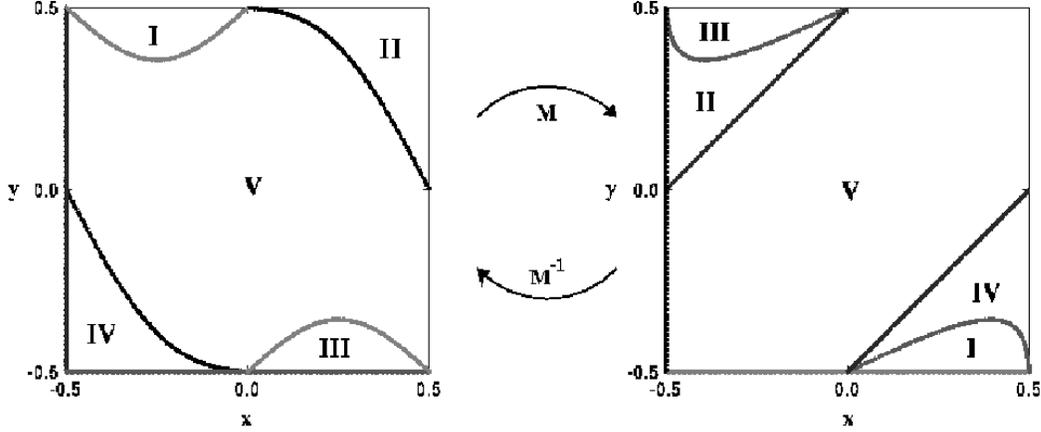}} 
\caption {The discontinuity curves of the standard map $M$
(\ref{eq:SM}) divide the phase space in five continuous regions
(I, II, III, IV, V). In each region the computation of the TD can
be performed accurately.} \label{MM-1}
\end{figure}

\begin{figure}
\centerline{ \epsfxsize=14.0cm \epsfbox{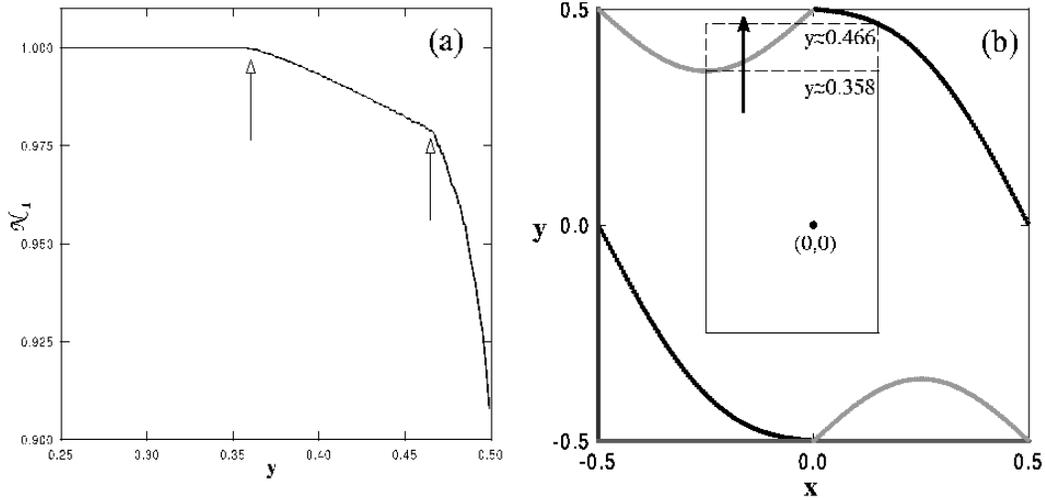}} 
\caption {(a) Number of period 1 fixed points $\cal{N}$$_1$ evaluated for the
SM (\ref{eq:SM}) with $k=0.9$ using the Kronecker integral
(\ref{eq:N2}), in a rectangular domain whose upper--side moves, as a
function of the $y$ coordinate of this side. The
rectangle and the discontinuity lines are shown in (b). For the
various rectangles, $\cal{N}$$_1$  should be equal to 1 since they
contain only 1 fixed point of period 1, namely point (0,0). The two
points marked by arrows in (a) where $\cal{N}$$_1$  deviates from the
correct value $\cal{N}$$_1 =1$, correspond to $y \approx 0.358$ and
$y \approx 0.466$ respectively, where the upper--side of the
rectangular crosses the two discontinuity curves in (b).}
\label{Fcross}
\end{figure}

For the SM the discontinuity curves are the lines $x=-0.5$ and
$y=-0.5$. By applying the SM map $M$ once these
lines are mapped on the curves seen in the right
panel of figure \ref{MM-1}. On the initial phase space there exist
also the discontinuity curves that will be mapped after one
iteration to the lines $x=-0.5$ and $y=-0.5$. These curves are
 also 
plotted in the left panel of figure
\ref{MM-1}. These curves can be produced by applying the inverse
SM to the discontinuity lines $x=-0.5$ and $y=-0.5$. So the
discontinuity curves divide the initial phase space in five
continuous regions marked as I, II, III, IV and V in figure
\ref{MM-1}. In each region the computation of the TD can be
performed accurately by Stenger's method or by evaluating Kronecker
integral. If, however, the boundary of the domain where
these procedures are applied, cross a discontinuity curve the
results we get are not correct (figure \ref{Fcross}).


In order to study the dependence of the procedure for finding the
TD in a region $D$, with respect to the distance of a root from the
boundary of $D$, we consider  the simple map
\begin{equation}
F^*= (f_1,f_2) = \left\{
\begin{array}{lcl}
f_1(x,y) & = & y-\frac{x^3}{3}+x \\
f_2(x,y) & = & y
\end{array}
\right. .\label{eq:F*}
\end{equation}
\begin{figure} 
\centerline{\epsfxsize=14.0cm \epsfbox{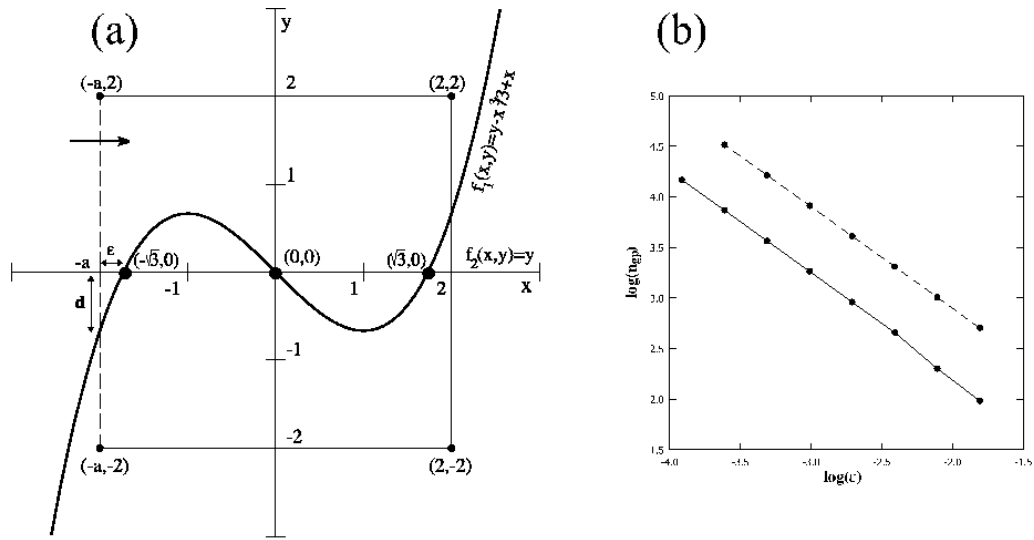}} 
\caption {(a) Plot of the curves $ f_1  = 
y-\frac{x^3}{3}+x=0$, $f_2= y=0$  (b) Dependence of the number of 
grid points $n_{gp}$, needed for computing the correct value 
of the TD in a domain, on the distance $\epsilon$ of a root from 
the boundary of the domain, for the set of equations of (a) 
(dashed line) and the SM (continuous line). } \label{test} 
\end{figure} 
The lines $f_1=0$, $f_2=0$ are plotted in figure \ref{test}(a).
The  system $F^*=(0,0)$ has three roots. The determinant of
the corresponding Jacobian matrix (det$J_{F^*}$) is positive for
root $(0,0)$ and negative for roots $(-\sqrt{3},0)$ and
$(\sqrt{3},0)$. We also consider a rectangular 
of the form $[-\mbox{a},2]\times [-2,2]$
with $\mbox{a}>\sqrt{3}$, shown in  figure \ref{test}(a). Since this
domain contains the three roots of the system the value of the TD is
$-1$. We set  $\mbox{a}=\sqrt{3}+\epsilon$ with $\epsilon >0$ so that the
boundary approaches the root as $\epsilon \to 0$, as shown by the
arrow in figure \ref{test}(a). We compute the TD for different
values of $\epsilon$ applying Stenger's method, by using the same number of
points $m$ on every side of the rectangle. We denote by
$n_{gp}=4m$ the smallest number of grid points needed to compute
the TD with certainty. In figure \ref{test}(b) we plot in log-log
scale, $n_{gp}$ with respect to $\epsilon$ (dashed line). The
slope of the curve is almost $-1$ so that $m \propto \epsilon^{-1}$.
The same result holds for any map when the
boundary approaches a root 
(the solid line in figure \ref{test}(b) is obtained for a
similar example for the SM (\ref{eq:SM})).

\begin{figure} 
\epsfxsize=14.5cm \epsfbox{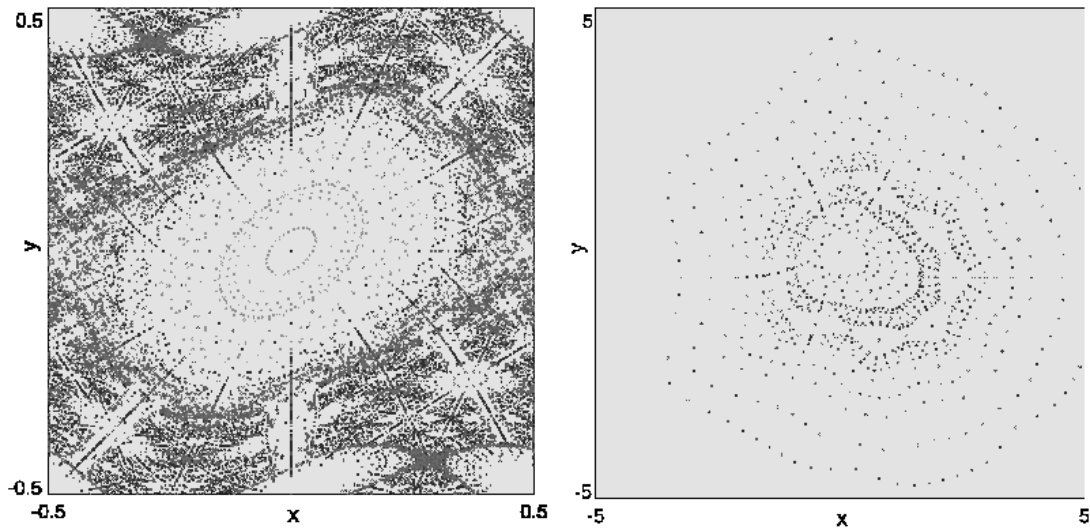}
\caption {Periodic orbits up to period $p=40$ for the SM 
(\ref{eq:SM}) for $k=0.9$ (left panel) and for the BM (\ref{eq:BM}) for 
$\omega=0.14$ (right panel). Different gray--scales correspond to 
periodic orbits with different kind 
of stability.} \label{POSM} 
\end{figure} 

Despite the problems caused by the discontinuity curves or by roots
located very close to the domain's boundary, the use of the characteristic
bisection method can locate a big fraction of the real periodic
orbits. Actually by applying the characteristic bisection method on the 
cells of a lattice formed by $2000 \times 2000$ grid points we were 
able to compute
a sufficient number of the periodic orbits with period up to 40,
for the SM (figure \ref{POSM}, left panel) and the BM (figure \ref{POSM}, 
right panel). The 
computed periodic orbits give us a clear picture of the dynamics of these maps.

\section{Synopsis}
We have studied the applicability of various numerical methods,
based on the topological degree theory, for locating high period
periodic orbits of 2D area preserving maps.

In particular we have used the Kronecker integral and applied the
Stenger's method for finding the TD in a bounded region of the
phase space. If the TD has a non-zero value we know that there
exist at least one periodic orbit in the corresponding region. The
computation of the TD for an appropriate set of equations allows
us to find the exact number of periodic orbits. We also applied
the characteristic bisection method on a mesh in the phase space
for locating the various fixed points.

The main advantage of all these methods is that they are not
affected by accuracy problems in computing the exact values of the
various functions used, as, the only computable information needed
is the algebraic signs of these values.

We have applied the above--mentioned methods to 2D symplectic
maps defined on $\RR^2$ and on the torus. The methods for
computing the TD are applied to continuous regions of the phase
space, so their use for maps on the torus is limited to regions
where no discontinuity curves exist. On the other hand the
characteristic bisection method proved to be very efficient for
all different types of maps, as, it allowed us to compute a
big amount of the real fixed points of period up to 40 in
reasonable computational times.

\section*{Acknowledgments}
Ch.~Skokos thanks the LOC of the conference for its financial
support. Ch.~Skokos was also supported by `Karatheodory'
post--doctoral fellowship No 2794 of the University of Patras and
by the Research Committee of the Academy of Athens (program No
200/532).

\end{document}